\documentclass[aps,prl,twocolumn,showpacs,superscriptaddress,floatfix,10pt]{revtex4-1}
\usepackage{lineno}
\usepackage{graphicx}
\usepackage{dcolumn}
\usepackage{bm}
\usepackage{amsfonts}
\usepackage{amssymb}
\usepackage{amsmath}
\usepackage{bbold}
\usepackage{fix-cm}
\usepackage{mathptmx} 
\usepackage[T1]{fontenc}
\usepackage[colorlinks,allcolors=blue]{hyperref}
\usepackage{color}

\newcommand{\al}{\alpha}

\newcommand{\az}{\varphi}

\newcommand{\oeq}{\begin{equation}}
\newcommand{\ceq}{\end{equation}}
\newcommand{\oeqn}{\begin{eqnarray}}
\newcommand{\ceqn}{\end{eqnarray}}

\renewcommand{\>}{\rangle}
\newcommand{\<}{\langle}

\newcommand{\sdf}{\,\,}

\renewcommand{\d}{{\mbox d}}

\newcommand{\hb}{\hbar}

\renewcommand{\vr}{{\bf r}}

\newcommand{\vR}{{\bf R}}

\newcommand{\mH}{{\mathcal{H}}}

\begin{document}

\title{How the Pauli exclusion principle affects fusion of atomic nuclei}

\author{C. Simenel}\email{cedric.simenel@anu.edu.au}
\affiliation{Department of Nuclear Physics, Research School of Physics and Engineering, The Australian National University, Canberra ACT  2601, Australia}

\author{A.S. Umar}\email{umar@compsci.cas.vanderbilt.edu}
\affiliation{Department of Physics and Astronomy, Vanderbilt University, Nashville, TN 37235, USA}

\author{K. Godbey}\email{kyle.s.godbey@vanderbilt.edu}
\affiliation{Department of Physics and Astronomy, Vanderbilt University, Nashville, TN 37235, USA}

\author{M. Dasgupta}
\affiliation{Department of Nuclear Physics, Research School of Physics and Engineering, The Australian National University, Canberra ACT  2601, Australia}

\author{D. J. Hinde}
\affiliation{Department of Nuclear Physics, Research School of Physics and Engineering, The Australian National University, Canberra ACT  2601, Australia}

\date{\today}

\begin{abstract}
The Pauli exclusion principle induces a repulsion between composite systems of identical fermions such as colliding atomic nuclei.
Our goal is to study how heavy-ion fusion is impacted by this ``Pauli repulsion''.
We propose a new microscopic approach, the density-constrained frozen Hartree-Fock method, to compute the bare potential including the Pauli exclusion principle exactly.
Pauli repulsion is shown to be important inside the barrier radius and increases with the charge product of the nuclei. 
Its main effect is to reduce tunnelling probability. 
Pauli repulsion is part of the solution to the long-standing deep sub-barrier fusion hindrance problem.
 \end{abstract}

\maketitle

The idea that identical fermions cannot occupy the same quantum state was proposed by Stoner\,\cite{stoner1924} and  generalized by Pauli\,\cite{pauli1925}.
Known as the Pauli exclusion principle, it was at first empirical, but is now  explained by the spin-statistic theorem in quantum field theory\,\cite{fierz1939,pauli1940}.
The importance of the ``Pauli exclusion principle'' cannot be overstated.
For instance, it is largely responsible for the stability of matter against collapse, as demonstrated by the existence of white dwarfs.
It also generates a repulsion between composite systems of identical fermions at short distance.
For example, it repels atomic electron clouds in ionic molecules due to the fermionic nature of the electron.
Another example is the hard-core repulsion between two nucleons induced by identical quarks of the same color present in both nucleons.
Naturally, a similar effect is expected to occur between atomic nuclei which are composite systems of nucleons.
Indeed, it has been predicted that the Pauli exclusion principle should induce a repulsion (called ``Pauli repulsion'' hereafter) between strongly overlapping nuclei\,\cite{fliessbach1971}.

The Pauli repulsion should then be included in the nucleus-nucleus potential used
to model reactions such as (in)elastic scattering, (multi)nucleon transfer, and fusion.
However, Pauli repulsion is usually neglected in these models:
it has been argued that the outcome of a collision between nuclei is mostly determined
at a  distance where the nuclei do not overlap much and thus the effects of the Pauli exclusion principle are minimized.
This argument is based on the assumption that nuclei do not necessarily probe
the inner part of the fusion barrier.
However, at energies well above the barrier, the system could reach more compact shapes
where one cannot neglect the effect of the Pauli principle anymore,
as was shown by several authors in the 1970's\,\cite{fliessbach1971,brink1975,zint1975,beck1978,sinha1979}.
Similarly, for deep sub-barrier energies the inner turning-point of the fusion barrier entails significant overlap
between the two nuclei\,\cite{dasso2003,umar2012a}.

Using a realistic microscopic approach to compute nucleus-nucleus bare potentials,
we  show that, in fact, the Pauli repulsion plays an important role on fusion at deep sub-barrier energies.
In particular, it provides a natural (though only partial) explanation for the experimentally observed deep sub-barrier fusion hindrance
\cite{jiang2002,dasgupta2007,stefanini2010}   (see Ref. \,\cite{back2014} for a review)
which has led to various theoretical interpretations
\cite{misicu2006,misicu2007,dasgupta2007,diaz-torres2008,diaz-torres2010,ichikawa2009b,ichikawa2015},
although none of them directly consider Pauli repulsion as a possible mechanism.

In order to investigate the effect of Pauli repulsion on heavy-ion fusion,
we introduce a novel microscopic method called density-constrained frozen Hartree-Fock (DCFHF)
to compute the interaction between nuclei while accounting exactly for the Pauli exclusion principle between nucleons.
The microscopically derived bare nucleus-nucleus potential including Pauli repulsion is then used to study deep sub-barrier fusion.
For simplicity, we focus on systems with doubly-magic nuclei which are spherical and non-superfluid. 
As an example, $^{40}$Ca$+^{40}$Ca, $^{48}$Ca$+^{48}$Ca and $^{16}$O$+^{208}$Pb reactions are studied theoretically and compared with experimental data.

To avoid the introduction of new parameters, we adopt the idea of Brueckner \textit{et al.}\,\cite{brueckner1968}
to derive the bare potential from an energy density functional (EDF)  $E[\rho]$
written as an integral of an energy density $\mH[\rho(\vr)]$, i.e.,
\oeq
E[\rho]=\int \d\vr \sdf \mH[\rho(\vr)]\,.
\ceq
The bare potential is obtained by requiring frozen ground-state densities $\rho_{i}$ of each nucleus ($i=1,2$) which we compute
using the Hartree-Fock (HF) mean-field approximation\,\cite{hartree1928,fock1930}. 
The Skyrme EDF\,\cite{skyrme1956} is used both in HF calculations and to compute the bare potential.
It accounts for the bulk properties of nuclear matter such as its incompressibility
which is crucial at short distances\,\cite{brueckner1968,misicu2006,hossain2015}.
Neglecting the Pauli exclusion principle between nucleons in different nuclei
leads to the usual frozen Hartree-Fock (FHF) potential\,\cite{denisov2002,washiyama2008,simenel2008,simenel2012}
\oeq
V_{FHF}(\vR)=\int \d\vr \sdf \mH[\rho_1(\vr)+\rho_2(\vr-\vR)] - E[\rho_1] -E[\rho_2],
\label{eq:frozen}
\ceq
where $\vR$ is the distance vector between the centres of mass of the nuclei.
The FHF potential, assumed to be central, can then directly be used to compute
fusion cross-sections\,\cite{simenel2013b,bourgin2016,vophuoc2016}.

Our new DCFHF method is the static counter-part of the density-constrained time-dependent Hartree-Fock
approach developed to extract the nucleus-nucleus potential of dynamically evolving systems\,\cite{umar2006b}.
In particular, this approach shows that the Pauli exclusion principle splits orbitals such that some
 states contribute attractively (bounding) and some repulsively (antibounding) to the potential\,\cite{umar2012b}.
In order to disentangle effects of the Pauli exclusion principle from the dynamics, we need to investigate
the bare potential without polarisation effects.
The dynamics can be included in a second step via, e.g., coupled-channels calculations \cite{simenel2013b}.

In the present method, it is important that the nuclear densities remain frozen as the densities
of the HF ground-states of the collision partners.
Consequently, the DCFHF approach facilitates the computation of the bare potential by using the self-consistent HF mean-field with exact frozen densities.
The Pauli exclusion principle is included exactly by allowing the single-particle states, comprising the combined nuclear density, to reorganize
to attain their minimum energy configuration and be properly antisymmetrized as the many-body
state is a Slater determinant of all the occupied single-particle wave-functions.
The HF minimization of the combined system is thus performed subject to the constraint that the
local proton ($p$) and neutron ($n$) densities do not change:
\begin{equation}
\delta \< \ H - \sum_{q=p,n}\int d\vr \ \lambda_q(\vr) \ [\rho_{1_q}(\vr)+\rho_{2_q}(\vr-\vR)] \ \> = 0\,,
\label{eq:var_dens}
\end{equation}
where the $\lambda_{n,p}(\vr)$ are Lagrange parameters at each point of space constraining the neutron and proton densities.
This equation determines the state vector (Slater determinant) $|\Phi(\vR)\>$.
The DCFHF potential, assumed to be central, is then defined as
\begin{equation}
V_{\mathrm{DCFHF}}(R)=\<\Phi(\vR) | H | \Phi(\vR) \>-E[\rho_1]-E[\rho_2]\,.
\label{eq:vr}
\end{equation}

\begin{figure*}[!htb]
\includegraphics*[width=18cm]{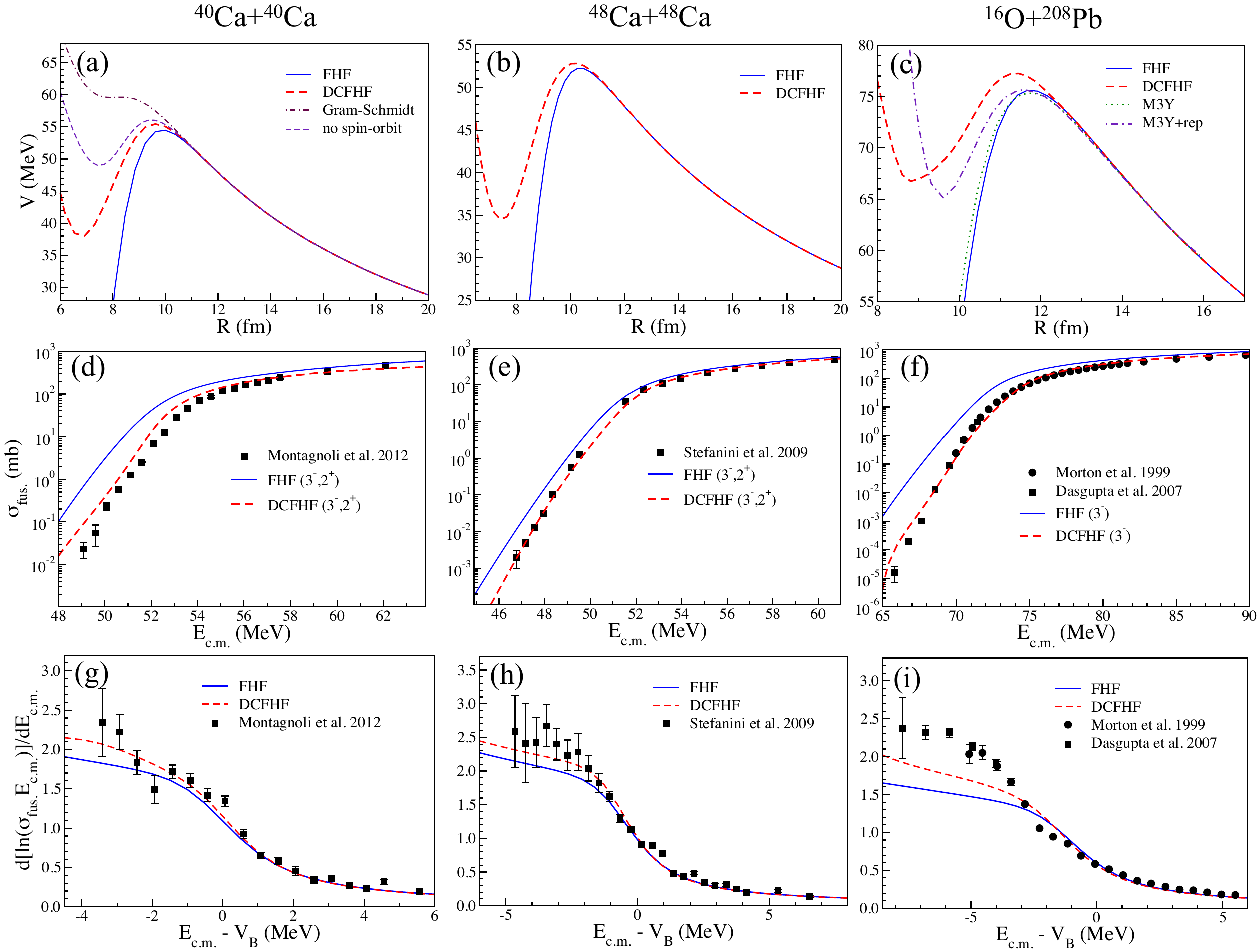}
\caption{(Color online) (a-c) Nucleus-nucleus potential without (FHF) and with (DCFHF) Pauli exclusion principle between nucleons of different nuclei. Potentials  from a Gram-Schmidt antisymmetrization (dotted-dashed line) and from DCFHF without rearrangement of the spin-orbit density (thin dashed line) are  shown in panel (a). M3Y (dotted line) and M3Y+rep (dotted-dashed line) phenomenological potentials \cite{esbensen2007}  are shown in panel (c). (d-f) Experimental\,\cite{morton1999,dasgupta2007,montagnoli2012,stefanini2009} and theoretical (with couplings to low-lying collective $2^+$ and/or $3^-$ states) fusion cross-sections $\sigma_{fus.}$ versus centre of mass energy $E_{c.m.}$. (d-f)  Logarithmic slopes of $\sigma_{fus.}E_{c.m.}$ versus $E_{c.m.}-V_B$ where $V_B$ is the barrier energy. In (g-i), FHF and DCFHF cross-sections are obtained without couplings, the latter being included via a shift in $E_{c.m.}$ (see text).}
\label{fig:Vb_cs}
\end{figure*}

 FHF and DCFHF calculations of bare nucleus-nucleus potentials were done in a three-dimensional
Cartesian geometry with no symmetry assumptions using a static version of the code of Ref.~\cite{umar2006c} and using the
Skyrme SLy4d interaction\,\cite{kim1997} which has been successful in describing various types of nuclear reactions\,\cite{simenel2012}.
The three-dimensional Poisson equation for the Coulomb potential
is solved by using Fast-Fourier Transform techniques
and the Slater approximation is used for the Coulomb exchange term.
The static HF equations and the DCFHF
minimizations are implemented using the damped gradient
iteration method. The box size used for all the calculations
was chosen to be $60\times 30\times 30$~fm$^3$, with a mesh spacing of
$1.0$~fm in all directions. 
These values provide very accurate results due to the employment of sophisticated discretization techniques\,\cite{umar1991a,umar1991b}.

The FHF (solid line) and DCFHF (dashed line) potentials are shown in Figs.~\ref{fig:Vb_cs}(a-c)
for $^{40}$Ca$+^{40}$Ca,   $^{48}$Ca$+^{48}$Ca, and $^{16}$O$+^{208}$Pb systems, respectively.
We observe that the Pauli exclusion principle (present only in DCFHF) induces a repulsion at short distance in the three systems.
The resulting effects are negligible outside the barrier and relatively modest near the barrier.
However, the impact is  more important in the inner barrier region, with the production of a potential pocket at short distance.
In $^{16}$O$+^{16}$O (not shown in Figs.~\ref{fig:Vb_cs}), the pocket height is negative and Pauli repulsion is expected to have a small impact on fusion in this system, except potentially at astrophysical energies. 
However, the pocket becomes shallower with increasing charge product $Z_1Z_2$. 
This is consistent with the fact that more nuclear overlap (and thus a larger Pauli repulsion) is required to compensate the larger Coulomb repulsion between the fragments. 
The most important effect of Pauli repulsion is to increase the barrier width. 
It is then expected to reduce the sub-barrier tunneling probability as the latter decreases exponentially with the barrier width.

In principle, the Pauli repulsion is expected to be energy dependent.
One source of energy dependence is the diminishing of the overlap between wave functions with relative kinetic momentum at higher
energies
reducing the Pauli repulsion\,\cite{fliessbach1971,brink1975,goritz1976,beck1978}.
Other sources are the dependence of the EDF on the current density (needed for Galilean invariance)\,\cite{brink1975} and
non-local effects of the Pauli exclusion principle leading to an energy dependence of the local equivalent potential\,\cite{schmid1982,chamon2002}.
These effects, however, are expected to impact the Pauli repulsion at energies much higher than the barrier (at least twice the barrier energy in $^{16}$O$+^{16}$O \cite{fliessbach1971,brink1975}),
and can then be neglected in near barrier fusion studies.

We have also tested other methods to account for Pauli repulsion in the bare potential.
For instance, antisymmetrizing overlapping ground-state wave-functions
\cite{fliessbach1971,brink1975,zint1975} can be done with a Gram-Schmidt procedure.
Although the resulting potential properly accounts for the Pauli exclusion principle,
it leads to much higher repulsion as illustrated in Fig.~\ref{fig:Vb_cs}(a) (dotted-dashed line)
for the $^{40}$Ca$+^{40}$Ca system in which the potential pocket, and therefore the fusion barrier, simply disappear.
Let us use a simple model to explain the origin of this large repulsion.
Consider two single-particle wave functions $\az_{1,2}$ belonging to the HF ground-states of
the two different nuclei and which have a small overlap in the neck region at $\vr_0$ only:
$\az_1^*(\vr)\az_2(\vr)\simeq \alpha \delta(\vr-\vr_0)$.
By definition, the total frozen density of these two nucleons is $\rho_F=|\az_1|^2+|\az_2|^2$.
The evaluation of observables, however, requires antisymmetrized wave-functions such as
$\tilde{\az}_\pm=\mathcal{N}_{\pm}(\az_1\pm\az_2)$ with  normalization coefficients
$\mathcal{N}_\pm=(2\pm\al\pm\al^*)^{-1/2}$ and overlaps $\<\tilde\az_-|\tilde\az_+\>=0$.
The corresponding density reads
$$\tilde\rho=|\tilde\az_+|^2+|\tilde\az_-|^2\simeq\rho_F-\frac{1}{2}(\al+\al^*)^2\delta(\vr-\vr_0).$$
It is reduced in the neck compared to the frozen density and thus leads to a smaller nuclear attraction
between the nuclei or, equivalently, to a spurious repulsion between the fragments as seen in Fig.~\ref{fig:Vb_cs}(a).
Naive antisymmetrization procedures are then not compatible with the frozen density picture.
This was also recognized in the earlier work concerning $\alpha$-nucleus scattering studies \,\cite{fliessbach1975},
where specialized normalization operators were developed to reconstruct the states following a
Gram-Schmidt orthogonalization. However, these methods could only be applied using semi-analytic
methods. The DCFHF achieves this without any approximation.

Let us now discuss another traditional method which is to account for Pauli repulsion
simply by increasing the kinetic energy density $\tau(\vr)$ (e.g., via the Thomas-Fermi model)
\cite{brink1975,zint1975,beck1978,denisov2010,nesterov2013}.
This method would be valid if the effect of the Pauli exclusion principle was only
to rearrange the kinetic energy term $\frac{\hb^2}{2m}\tau$ without impacting other terms of the functional.
In fact, the EDF also depends on $\tau$ via the ``$t_{1,2}$'' momentum dependent terms of the Skyrme  effective interaction
and, then, a variation of $\tau(\vr)$ also affects the nuclear part of the potential\,\cite{brink1975,denisov2010}.
At the same time, we have also observed that including the Pauli exclusion principle has a strong impact on the spin-orbit energy.
This is illustrated in Fig.~\ref{fig:Vb_cs}(a) for the $^{40}$Ca$+^{40}$Ca system.
For this system, removing the spin-orbit interaction has little impact on the FHF potential (not shown in the figure), but strongly increases the repulsion between the fragments in the DCFHF potential (thin dashed line).
This shows that the spin-orbit energy absorbs a large part of the Pauli repulsion.
Thus, the Pauli exclusion principle has a more complicated effect than just increasing the kinetic energy density.

Coupled-channels calculations of fusion cross-sections were performed with the  \textsc{ccfull} code\,\cite{hagino1999}
using Woods-Saxon fits of the FHF and DCFHF potentials.
By default, the incoming wave boundary condition (IWBC) was used.
For shallow pocket potentials, however,
the IWBC should be replaced by an imaginary potential inside the barrier to avoid numerical instabilities.
This is done for calculations with the $^{16}$O$+^{208}$Pb DCFHF potential using a modified version of \textsc{ccfull}.
Couplings to the low-lying collective $2^+$ (in calcium isotopes) and $3^-$  states are included
with standard values of the coupling constants\,\cite{morton1999,rowley2010}.
One (two) vibrational mode(s) can be included in the projectile (target).
For the $2^+$ states, we then use the fact that, for symmetric systems,
the mutual excitation of one-phonon states in both nuclei can be approximated
by one phonon with a coupling constant scaled by $\sqrt{2}$\,\cite{esbensen1987}.
Here, the CC calculations are kept simple and include only the most relevant couplings.
Improvements could be obtained, e.g., by including anharmonicity of the multi-phonon states\,\cite{yao2016}.
The resulting fusion cross-sections are plotted in Figs.~\ref{fig:Vb_cs}(d-f). 
Calculations with the FHF potential systematically overestimate the data while the DCFHF
potential leads to a much better agreement with experiment at all energies, and ranging over eight orders of magnitude in cross-sections.
This shows the importance of taking into account Pauli repulsion in  the bare potential for fusion calculations.
We emphasise that these calculations are performed without adjustable parameters.

The behaviour of fusion at deep sub-barrier energies is often studied using  the  logarithmic slope 
$d(\sigma_{fus.}E_{c.m.})/dE_{c.m.}$. 
Large logarithmic slopes are a signature of a rapid decrease of $\sigma_{fus.}$ with decreasing energy. 
Deep-sub-barrier fusion hindrance is characterised by the failure of theoretical models to reproduce  large logarithmic slopes observed experimentally at low energy. 
To avoid numerical instabilities due to shallow potentials in the calculations of logarithmic slopes, couplings to internal excitations of the nuclei have been removed in the calculations of barrier transmission and accounted for via an overall lowering of $V_B$ by less than $5\%$ depending on the structure of the reactants \cite{dasgupta2007}. 
Indeed, it has been shown that couplings have little effects on the logarithmic slope at these energies \cite{dasgupta2007}. 

We see in Fig.~\ref{fig:Vb_cs}(g-i) that the inclusion of Pauli repulsion in DCFHF indeed increases the logarithmic slope at low energy. 
Although Pauli repulsion is shown to play a crucial role, it is not yet sufficient to reproduce  experimental data at deep sub-barrier energies. 
Other contributions are expected to come from dissipative effects \cite{dasgupta2007} and from the transition between the nucleus-nucleus potential to the one-nucleus adiabatic potential\,\cite{ichikawa2009b}.
However, repulsive effects  from the incompressibility of nuclear matter invoked in\,\cite{misicu2006} are not observed in our microscopic calculations. 
Both the FHF and DCFHF calculations use the same Skyrme functional (SLy4d) with a realistic compression modulus of the symmetric nuclear matter $K_\infty\simeq230$~MeV. 
Although the FHF potential properly takes into account  effects due to incompressibility, it is very close to standard phenomenological potentials. 
We illustrate this with the example of the M3Y potential\,\cite{misicu2006} in Fig.~\ref{fig:Vb_cs}(c). 
The addition of a repulsive component at short distance [M3Y+rep parametrisation shown with a dotted-dashed line Fig.~\ref{fig:Vb_cs}(c)],   introduced phenomenologically in\,\cite{misicu2006}  to explain experimental fusion data at deep sub-barrier energies, can then not be justified by an effect of incompressibility. 
It is more likely that it simulates other effects such as  Pauli repulsion. 

C. S. thanks E. C. Simpson for useful discussions.
This work has been supported by the
Australian Research Council Grant No. FT120100760,
and by the U.S. Department of Energy under grant No.
DE-SC0013847 with Vanderbilt University.

\bibliography{VU_bibtex_master}

\end{document}